# MAGNETO–SUPERCONDUCTIVITY IN RUTHENO-CUPRATES $RuSr_2GdCu_2O_{8-d}$ (Ru-1212) AND $RuSr_2(Gd_{0.75}Ce_{0.25})_2Cu_2O_{10-d}$ (Ru-1222): A CRITICAL REVIEW

**V.P.S. AWANA, M. KARPPINEN, AND H. YAMAUCHI**

*Materials and Structures Laboratory, Tokyo Institute of Technology, Yokohama 226-8503, JAPAN*

## 1. INTRODUCTION

According to a long-term common sense superconductivity and magnetic long-range order do not mutually exist within a single (thermodynamical) phase. The topic has been widely discussed in condensed matter physics over decades. Nevertheless, coexistence of high-$T_c$ superconductivity and magnetism was reported for a rutheno-cuprate of the Ru-1222 type, *i.e.* $RuSr_2(Gd_{0.7}Ce_{0.3})_2Cu_2O_{10-\delta}$ [1], and more recently for $RuSr_2GdCu_2O_{8-\delta}$ (Ru-1212) [2]. These reports have renewed the interest in the possible coexistence of superconductivity and magnetism. It is believed that in rutheno-cuprates the $RuO_6$ octahedra in the charge reservoir are mainly responsible both for magnetism and for doping holes into the superconductive $CuO_2$ plane.

The structures of both $RuSr_2GdCu_2O_{8-\delta}$ and $RuSr_2(Gd,Ce)_2Cu_2O_{10-\delta}$ are derived from that of $R$$Ba_2Cu_3O_{7-\delta}$ or $CuBa_2RCu_2O_{7-\delta}$, with Cu in the charge reservoir replaced by Ru such that the $CuO_{1-\delta}$ chain is replaced by a $RuO_{2-\delta}$ sheet [3]. In the $RuSr_2(Gd,Ce)_2Cu_2O_{10-\delta}$ structure furthermore, a three-layer fluorite-type block instead of a single oxygen-free $R$ (= rare earth element) layer is inserted between the two $CuO_2$ planes of the Cu-1212 structure. The $RuSr_2(Gd,Ce)_2Cu_2O_{10-\delta}$ phase with a layer sequence of SrO-$RuO_{2-\delta}$-SrO-$CuO_2$-(Gd,Ce)-$O_2$-(Gd,Ce)-$CuO_2$ is an $m$ = 1, $k$ = 1, $s$ = 2 member of "category-B" cuprates, $M_mA_{2k}B_sCu_{1+k}O_{m+4k+2s\pm\delta}$ or $M$-$m(2k)s(1+k)$, *i.e.* Ru-1222, while $RuSr_2GdCu_2O_{8-\delta}$ is an $m$ = 1, $r$ = 2, $n$ = 2 member of "category-A" cuprates, $M_mA_rQ_{n-1}Cu_nO_{m+2+2n\pm\delta}$ or $M$-$mr(n-1)n$, *i.e.*

Ru-1212 [4]. Note that according to this general categorization scheme, the standard $CuBa_2RCu_2O_{7-\delta}$ phase is referred to as Cu-1212.

Although bulk magnetism due to ordering of the Ru moments was confirmed for Ru-1212 from μSR (muon spin rotation/relaxation/resonance) and ESR (electron spin resonance) studies [2,5], the exact type of ordering is still debated. In particular, the results revealed from neutron scattering experiments [6-9] and magnetization studies [10-13] do not agree with each other. While the former concludes the ordering to be of antiferromagnetic nature, the latter indicates some ferromagnetic ordering. The appearance of bulk superconductivity at low temperatures was initially criticised [14]. However, a recent report argues that bulk superconductivity exists in this compound within a magnetically ordered state [15]. Appearance of bulk superconductivity in Ru-1212 is also confirmed from specific heat ($C_P$) measurements [16,17], though the existing reports do not agree with each other in terms of $C_P$ measurements under magnetic field [$C_p(H,T)$]. In particular, the superconducting transition temperature ($T_c$) as viewed from the $C_P$ peak increases with field in one report [16] but decreases with field in another [17]. While the former indicates towards the triplet pairing, the latter suggests a normal underdoped high-$T_c$ superconductor (HTSC) case. It is also suggested that small impurity of $GdSr_2RuO_6$ presumably present in the samples of Ref. 16 is responsible for the $C_P(H,T)$ behaviour being different from that of Ref. 17. Also worth noting is the fact that, though μSR and ESR studies indicate towards bulk nature of magnetic ordering [2,5], the $C_p$ measurements done by the same group of authors do not reveal a distinct peak at the ordering temperature [16]. A hump in $C_p$ seen at the magnetic ordering temperature is rather indicative of either short-range magnetic correlations or low-dimensional magnetic ordering of Ru spins.

Even though "magneto-superconductivity" was first realised in Ru-1222, a lot still remains about the physical characterization of the phase. However the main features are the same for both Ru-1212 and Ru-1222. The magnetic structure of Ru-1222 has been studied by neutron powder diffraction [18]. Despite the fact that various physical-property measurements have been carried out on Ru-1212 [2,6-17] and Ru-1222 [1,18-23], no final consensus has been reached, *i.e.* discussion on their basic characteristics in terms of the oxygen stoichiometry, valence state of Ru, carrier concentration and doping mechanism has not been completed yet. This becomes more important in the event when contradictory experimental results are obtained on different samples [2,6,8,10,11,14,17,21-23]. Also, it has been reported [24] that solid solutions of composition $(Ru_{1-x}Cu_x)Sr_2GdCu_2O_{8-\delta}$ can form within $0 < x < 0.75$ with $T_c$ up to 74 K. Interestingly with the higher $x$ values in the above composition the Ru spins do not order magnetically down to 5 K. Henceforth to conclude the coexistence of long-range magnetic ordering of Ru spins with superconductivity in the $CuO_2$ plane, one should strictly avoid the formation of $(Ru_{1-x}Cu_x)$-1212 solid solutions in pristine Ru-1212. Worth mentioning is the fact that Ru/Cu intermixing becomes more complicated as the two elements cannot be distinguished without ambiguity by neutron diffraction, a technique commonly used for fixing various cation occupancies in inorganic solids. Both Ru and Cu do have nearly the same scattering cross-sections for thermal neutrons.

Here it is important to note that the possible coexistence of superconductivity and magnetism (in particular ferromagnetism) has been a topic of debate for decades even in the case of more homogeneous intermetallic compounds. (For a recent update of the topic, see *http://physicsweb.org/article/world/15/19*.) In particular the microscopic length scales

in terms of the superconducting coherence length and the long-range magnetic ordering are important. Also in HTSC compounds, large anisotropy within the unit cell and microscopic compositional variations due to different cation intermixing or the variation in oxygen content are often seen. In such a situation the possibility of magnetic but not superconducting solid solutions coexisting microscopically with superconducting but not magnetic material cannot be ruled out without ambiguity. Such a possibility was indicated in the very beginning of the research on the "possible coexistence of superconductivity and magnetism in rutheno-cuprates" [11,14,25]. The concern of phase purity at the microscopic level in both Ru-1212 and Ru-1222 still remains unresolved. Also, we should look more carefully at the existing contradictions in the reported literature on rutheno-cuprates. Nevertheless, results of recent NMR (nuclear magnetic resonance) experiments on Ru-1212 were interpreted in terms of the coexistence of superconductivity and magnetism [26].

In the current review, we not only access the existing literature critically, but also present our own data in terms of phase formation and structural, thermal, magnetic, electrical, spectroscopic and microscopic characterization for both Ru-1212 and Ru-1222. We further conclude that the coexistence of superconductivity and magnetism, in particular the ferromagnetism at microscopic level in both these phases is far from conclusive.

## 2. EXPERIMENTAL DETAILS

Samples of $RuSr_2GdCu_2O_{8-\delta}$ and $RuSr_2(Gd_{0.75}Ce_{0.25})_2Cu_2O_{10-\delta}$ were synthesized through a solid-state reaction route from stoichiometric amounts of $RuO_2$, $SrO_2$, $Gd_2O_3$, $CeO_2$ and CuO. Calcinations were carried out on mixed powders at 1000 $^oC$, 1020 $^oC$ and 1040 $^oC$ for 24 hours at each temperature with intermediate grindings. The pressed bar-shaped pellets were annealed in a flow of oxygen at 1075 $^oC$ for 40 hours and subsequently cooled slowly over a span of another 20 hours down to room temperature. These samples are termed as "as-synthesized". Part of the as-synthesized samples were further annealed in high-pressure oxygen (100 atm) at 420 $^oC$ for 100 hours and subsequently cooled slowly to room temperature. These samples are termed as "100-atm $O_2$-annealed". Though the heat treatments used for the samples in our study are in general similar to those as reported in literature [1,2,8,9,21-23], minor differences do exist from one laboratory to another in terms of annealing hours and the temperatures used. Also, it has been reported that not always all samples of the same batch with similar heating schedule show superconductivity [14,27]. Our general experience is also the same particularly for Ru-1212, in which achieving superconductivity seems to be a tricky job. Worth mentioning is the fact that for both Ru-1212 and Ru-1222, single-phase samples are achieved only for $R$ = Gd, Sm and Eu, with the normal heating schedules mentioned above. For $R$ = Y and Dy, *etc*., one needs to employ the HPHT (high-pressure high-temperature) procedure for attaining the Ru-1212 phase [7,13].

Thermogravimetric (TG) analyses (Perkin Elmer: System 7) were carried out in a 5 % $H_2$/95 % Ar atmosphere at the rate of 1 $^oC$/min to investigate the oxygen non-stoichiometry. X-ray diffraction (XRD) patterns were collected at room temperature (MAC Science: MXP18VAHF[22]; Cu$K_\alpha$ radiation). Magnetization measurements were carried out on a superconducting-quantum-interference-device (SQUID) magnetometer (Quantum Design:

MPMS-5S). Resistivity measurements under an applied magnetic field of 0 - 7 T were performed in the temperature range of 5 - 300 K using a physical-property-measurement system (Quantum Design: PPMS). Electron diffraction (ED) patterns were obtained using a transmission electron microscope (TEM; Hitachi H-9000) operated at an accelerating voltage of 300 kV. The Ru $L_{III}$-edge XANES measurements were performed at room temperature for polycrystalline samples at the BL15B beamline of the Synchrotron Radiation Research Center (SRRC) in Hsinchu, Taiwan.

## 3. RESULTS & DISCUSSION

### 3.1 Phase formation and lattice parameters: X-ray diffraction results

The presently studied $RuSr_2GdCu_2O_{8-\delta}$ samples possess a tetragonal Ru-1212 structure with a space group *P*4/*mmm*. For an as-synthesized sample lattice parameters were determined at $a = b = 3.8218(6)$ Å and $c = 11.476(1)$ Å. Corresponding X-ray diffraction pattern is shown in Fig. 1. Small amount of $SrRuO_3$ is also seen, which is marked on the pattern.

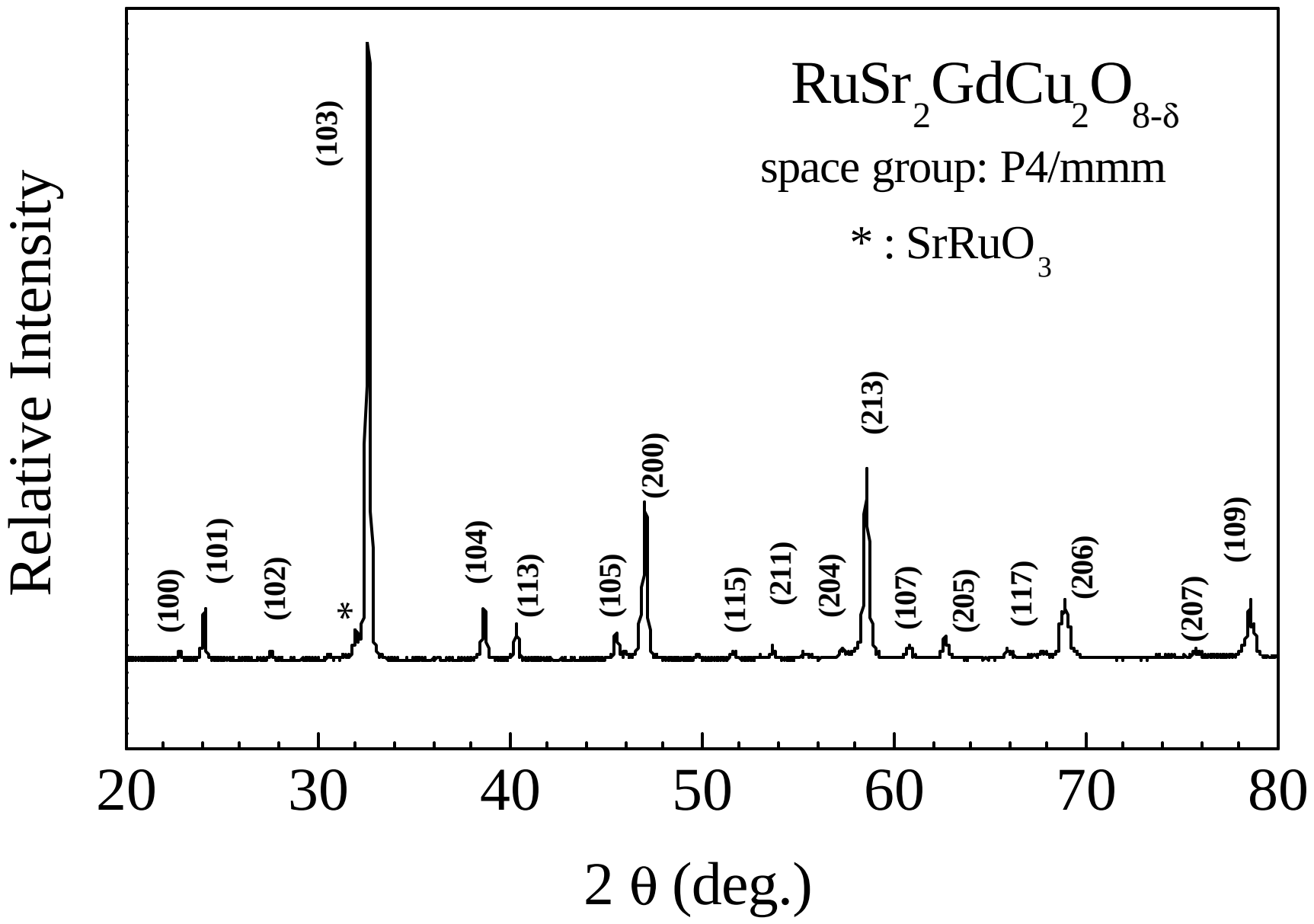


**Fig. 1.** X-ray diffraction pattern for an as-synthesized $RuSr_2GdCu_2O_{8-\delta}$ sample.

Essentially no difference was found in X-ray diffraction pattern or lattice parameters for the 100-atm $O_2$-annealed sample, suggesting that the oxygen content remained unchanged upon the high-$O_2$-pressure annealing. Worth mentioning is the fact that earlier one of us obtained a phase-pure (no trace of $SrRuO_3$ within the XRD detection limit) Ru-1212

sample that was not superconducting, even with various types of post-annealing treatments [27]. We will discuss this later.

As compared with Ru-1212, the Ru-1222 phase forms more easily in impurity-free form. Both the as-synthesized and the 100-atm $O_2$-annealed Ru-1222 samples were found to be of high quality in terms of phase purity. An X-ray diffraction pattern for an as-synthesized sample is shown in Fig. 2. The lattice parameters were determined from the diffraction data in the tetragonal space group *I4/mmm*: $a = b = 3.8337(6)$ Å and $c = 27.493(1)$ Å for the as-synthesized sample, and $a = b = 3.8327(7)$ Å and $c = 27.393(1)$ for the 100-atm $O_2$-annealed sample. The shorter lattice parameters for the 100-atm $O_2$-annealed sample are believed to manifest the fact that it is more completely oxygenated than the as-synthesized sample.

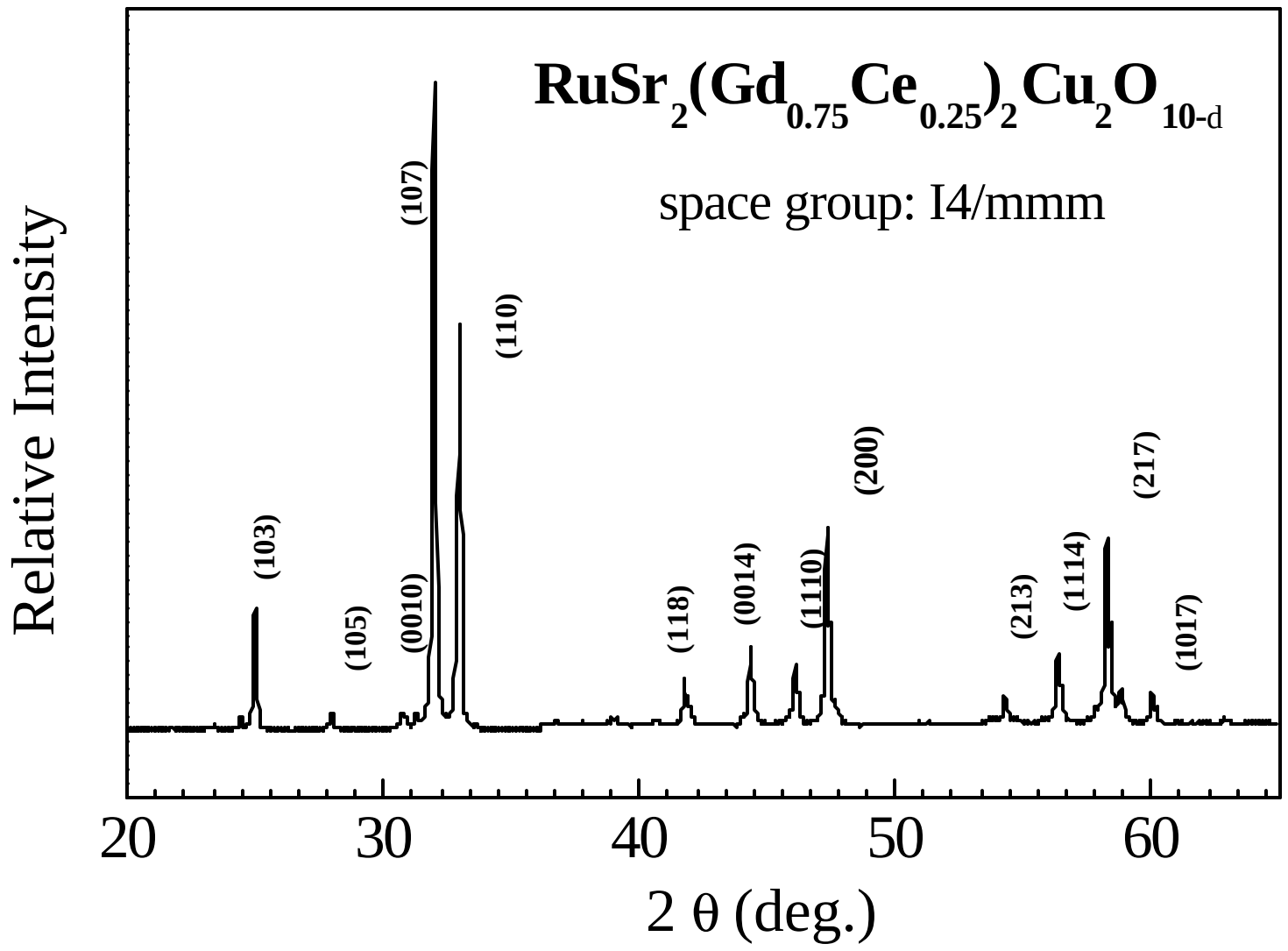


**Fig. 2.** X-ray diffraction pattern for an as-synthesized $RuSr_2(Gd_{0.75}Ce_{0.25})_2Cu_2O_{10-\delta}$ sample.

X-ray diffraction results, in terms of phase purity, space groups and the obtained lattice parameters are in general accordance with the reported literature [1,2,8,14,21-23].

### 3.2 Oxygen stoichiometry: TG results

Thermogravimetric reduction curves were recorded for both as-synthesized and 100-atm $O_2$-annealed samples of Ru-1212 and Ru-1222 [28]. Two examples of the obtained TG curves are shown in Fig. 3. Upon heating in flowing $H_2$/Ar gas both the phases, Ru-1212 and Ru-1222, release oxygen in a parallel manner decomposing to Ru metal, Cu metal and oxides of Sr, Gd and Ce in two distinct steps about 300 and 450 $^oC$. Owing to the sharpness of the weight loss behaviour such reductive decomposition carried out in a thermobalance

may be utilized in precise oxygen content determination for these phases. It was found that the two Ru-1212 samples, as-synthesized and 100-atm $O_2$-annealed, possess nearly the same oxygen content. For the Ru-1222 phase, several repeated experiments revealed that the difference in oxygen content between the as-synthesized and the 100-atm $O_2$-annealed samples is ~0.15 oxygen atoms *per* formula unit [28,29]. In terms of the oxygen-content variation, the TG results thus are consistent with the conclusions made based on the XRD data.

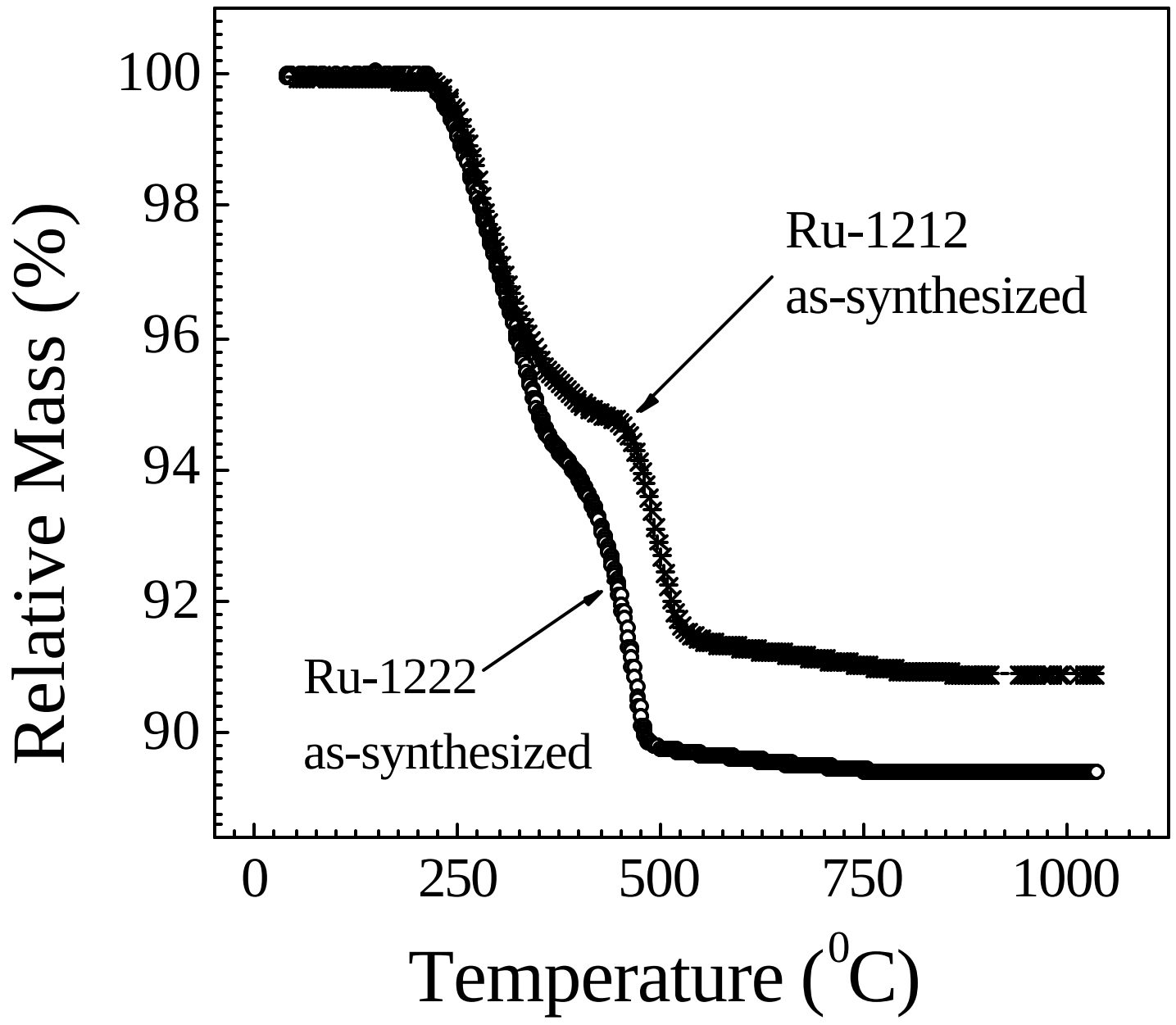


**Fig. 3.** TG curves for as-synthesized $RuSr_2GdCu_2O_{8-\delta}$ and $RuSr_2(Gd_{0.75}Ce_{0.25})_2Cu_2O_{10-\delta}$ samples recorded in 5 % $H_2$/95 % Ar atmosphere.

### 3.3 Valence of Ru: XANES spectroscopy results

The Ru $L_{III}$-edge XANES spectra obtained for an as-synthesized and a 100-atm $O_2$-annealed Ru-1222 sample are displayed in Fig. 4. The spectra were analyzed quantitatively by fitting them to certain linear combinations of those for reference materials $Sr_2RuO_4$ ($Ru^{IV}$) and $Sr_2GdRuO_6$ ($Ru^V$). All the spectra exhibited two peaks, the higher-energy one and the lower-energy one being due to $2p \rightarrow e_g$ and $2p \rightarrow t_{2g}$ transitions, respectively [30-32]: with increasing Ru valence from +IV to +V, the crystal-field splitting increases and thereby the separation between the two peaks enhances. Furthermore, the peaks are accordingly shifted by ~1.5 eV to the higher energy. From Fig. 4, both the Ru-1222 samples are between the two reference materials in terms of the Ru valence. Fitting the spectra revealed a valence value of +4.74 for the as-synthesized sample and +4.81 for the 100-atm $O_2$-annealed sample [29].

The obtained result suggests that the valence of Ru in Ru-1222 is affected by the change in oxygen content. It is therefore interesting to compare the presently obtained Ru valence values to that previously reported for a $RuSr_2(Gd_{0.7}Ce_{0.3})_2Cu_2O_{10-\delta}$ sample (+4.95) with $T_c \approx 60$ K [31]. For the three Ru-1222 samples the Ru valence/$T_c$ values thus were: +4.74/30 K, +4.81/43 K and +4.95/60 K. (Note that the XANES measurements and analyses were carried out in parallel ways for all the three samples.) The latter two samples were both annealed under 100 atm oxygen pressure, but with different temperature programs. It is thus likely that the one previously reported [31] had somewhat higher oxygen content than the present 100-atm $O_2$-annealed sample. It seems that the valence of Ru in Ru-1222 depends on the oxygen content, thus indirectly suggesting that the changes in oxygen stoichiometry occur in the $RuO_{2-\delta}$ layer. Here it is interesting to note that a previous study had shown that Ru remains essentially unchanged (close to pentavalent) upon varying the Ce-substitution level within $0.3 \leq x \leq 0.5$ in fully oxygen-loaded $RuSr_2(Gd_{1-x}Ce_x)_2Cu_2O_{10-\delta}$ samples [31].

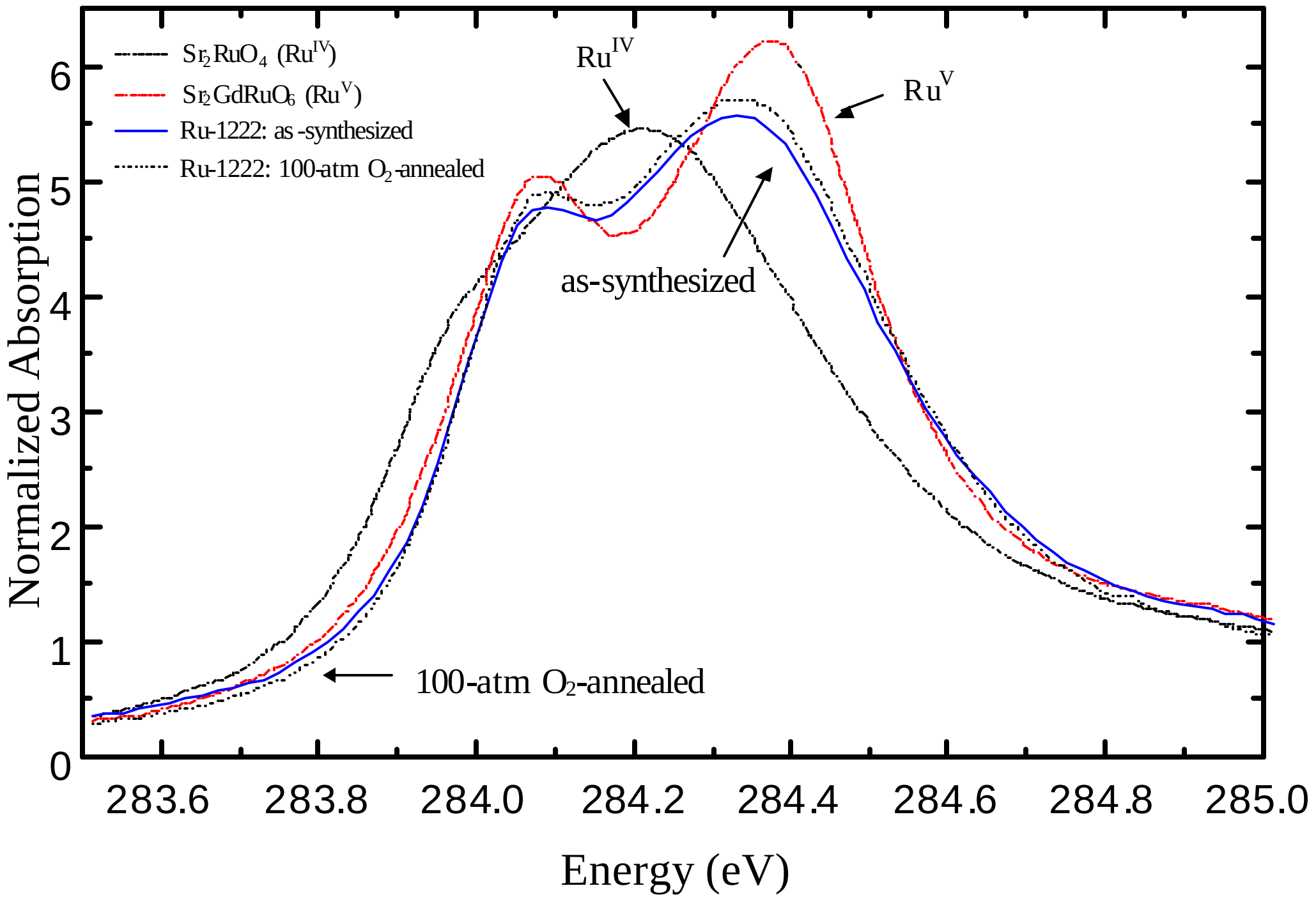


**Fig. 4.** Ru $L_{III}$-edge XANES spectra for reference materials, $Sr_2RuO_4$ ($Ru^{IV}$) and $Sr_2GdRuO_6$ ($Ru^V$), and for as-synthesized and 100-atm $O_2$-annealed $RuSr_2(Gd_{0.75}Ce_{0.25})_2Cu_2O_{10-\delta}$ samples.

Ru $L_{III}$-edge XANES analysis for an as-synthesized (final annealing in $O_2$ at 1075 °C) $RuSr_2GdCu_2O_{8-\delta}$ sample revealed a valence value of +4.62 [28]. This value is very close to that reported in Ref. [30] for a $RuSr_2GdCu_2O_{8-\delta}$ sample, *i.e.* +4.60, with the final annealing performed in $O_2$ at 1060 °C.

### 3.4 Superstructures: SAED results

**A. As-synthesized Ru-1212 sample**

Figure 5 depicts a selected-area electron diffraction (SAED) pattern of the *a-b* plane for a superconducting, as-synthesized $RuSr_2GdCu_2O_{8-\delta}$ sample. Two superstructures are seen: one weak spot at the centre of the *a-b* rectangle and the other along the *b* direction. Earlier microstructural study on samples of the Ru-1212 phase showed only one superstructure along the *a-b* plane, that is, the spot at the centre of the *a-b* rectangle. This was interpreted as being due to tilting of the $RuO_6$ octahedra [33]. For the sample presently studied showing the additional superspot along the *b* direction there is a possibility that either Ru/Cu or vacancy ordering of 2*b* periodicity is taking place along the *b* direction. It thus seems that superconducting clusters of composition $(Ru_{1-x}Cu_x)Sr_2GdCu_2O_{8-\delta}$ may be present within the $RuSr_2GdCu_2O_{8-\delta}$ phase giving rise to SNS/SIS (superconductor-normal metal-superconductor/superconductor-insulator-superconductor) junctions in the resulting material. This would explain the different type of broadening of resistive transition under a magnetic field for many Ru-1212 samples [10,11,14,17,25] along with the present one, see section 3.5. Though the possible presence of SNS/SIS junctions in various Ru-1212 samples could be strictly sample dependent, we argue that all reported Ru-1212 magneto-superconducting samples may not be homogenous in composition, but Ru/Cu ordering at the charge-reservoir cation site may be present in some of them. Also we would like to mention that we also get the *b*-direction clean *a-b* plane with only $RuO_6$ tilting superstructure in the same sample. Electron diffraction pattern from the *a-c* plane of the same sample is seen clean without any superstructures.

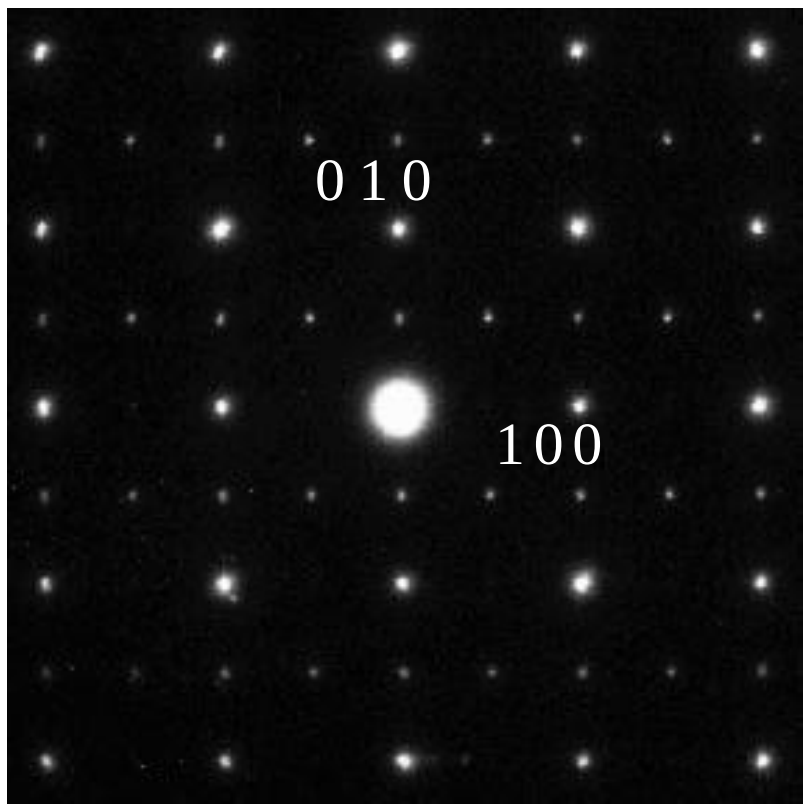


**Fig. 5.** SAED pattern for an as-synthesized Ru-1212 sample taken with incident beam // *c*.

**B. As-synthesized Ru-1222 sample**

Figure 6 shows a SAED pattern from the *a-c* plane of an as-synthesized $RuSr_2(Gd_{0.75}Ce_{0.25})_2Cu_2O_{10-\delta}$ sample. No superstructures are seen. The *c*-parameter length

calculated from the SAED pattern is around 27.5 Å, being close to the value determined from the X-ray diffraction data *i.e*. 27.493(1) Å.

The SAED pattern taken for the same sample from the *a-b* plane is shown in Fig. 7. Interestingly the *a-b* plane observed here is also clean without any superstructure. Recent neutron powder diffraction (NPD) studies [18] on Ru-1222 revealed that even though the $RuO_6$ tilt angle is seen, no evidence of supercell reflections arising from extended regions of $RuO_6$ rotations could be observed, in contrast to the case of the Ru-1212 phase [8,33]. Our SAED results thus corroborate the NPD results [18] in terms of the absence of superstructural spot at the centre of the *a-b* rectangle. It should however be remembered that the as-synthesized Ru-1222 sample studied here is deficient in oxygen.

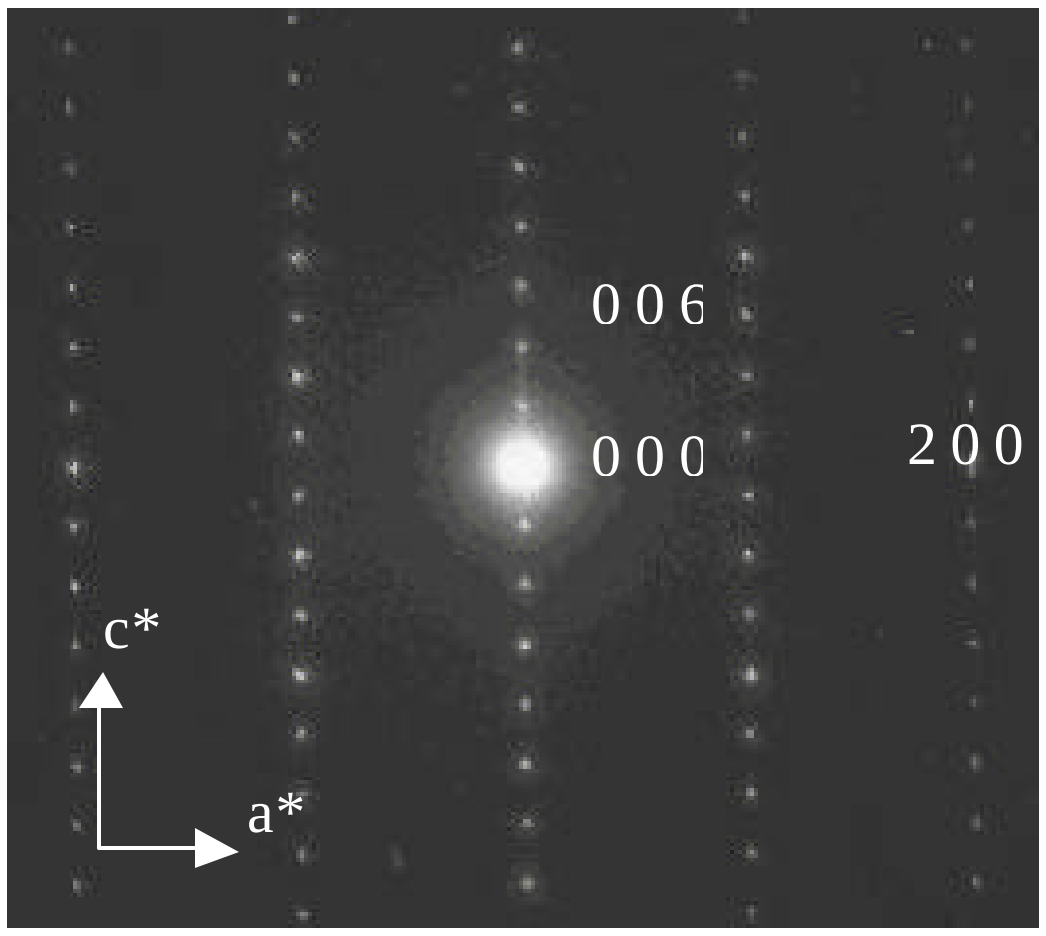


**Fig. 6.** SAED pattern for an as-synthesized Ru-1222 sample taken with incident beam // *b*.

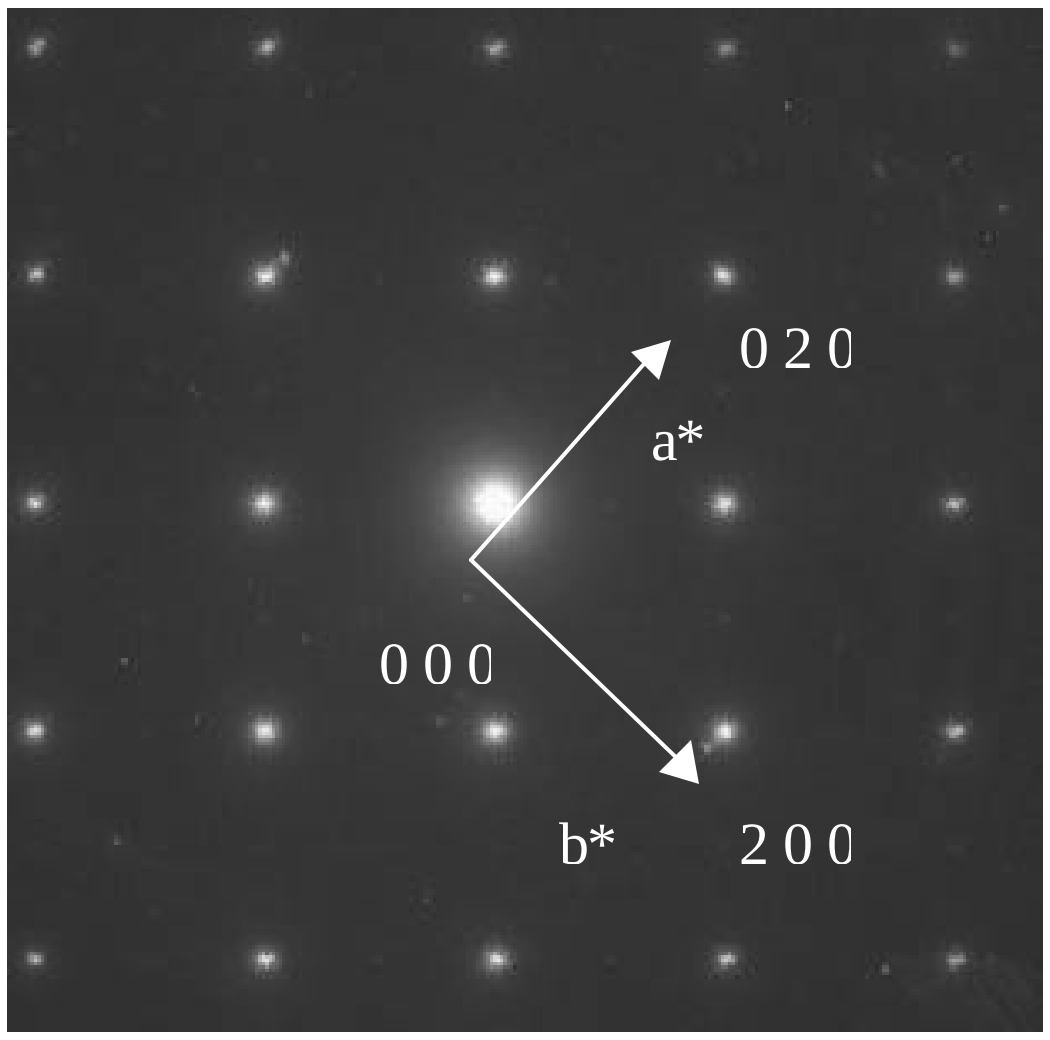


**Fig. 7.** SAED pattern for an as-synthesized Ru-1222 sample taken with incident beam // *c*.

### 3.5 Magneto-superconductivity and magnetic characteristics: SQUID results

#### A. Non-superconducting Ru-1212 samples

Figure 8 depicts both ZFC (zero-field-cooled) and FC (field-cooled) magnetic moment (*M*) *versus* temperature (*T*) plots for various $RuSr_2GdCu_2O_{8-\delta}$ samples with an applied field of 10 Oe [27]. As is seen from this figure the ZFC and FC magnetisation curves show a significant branching around 145 K for the as-synthesized sample with a further sharp drop in magnetisation around 2.6 K.

The branching of ZFC and FC curves at 145 K originates from the magnetic ordering of Ru moments and the sharp peak at 2.6 K is due to antiferromagnetic ordering of Gd moments. Interestingly in presently studied non-superconducting $RuSr_2GdCu_2O_{8-\delta}$ samples the ordering temperature of Ru moments appears to be 13 K higher than the value reported for similar but superconducting samples [1,6,8]. Both the ZFC and FC branching about 145 K and the sharp peak at 2.6 K are seen for all the samples including the high-$O_2$-pressure annealed and the argon annealed ones. None of the samples show any traces of superconductivity down to 2 K, even with very low field measurements at 1.5 Oe. Worth reminding is the fact that these samples of ours contain no traces of $SrRuO_3$. These non-superconducting $RuSr_2GdCu_2O_{8-\delta}$ samples showed a similar ferromagnetic component at 5 K, as for reported superconducting samples.

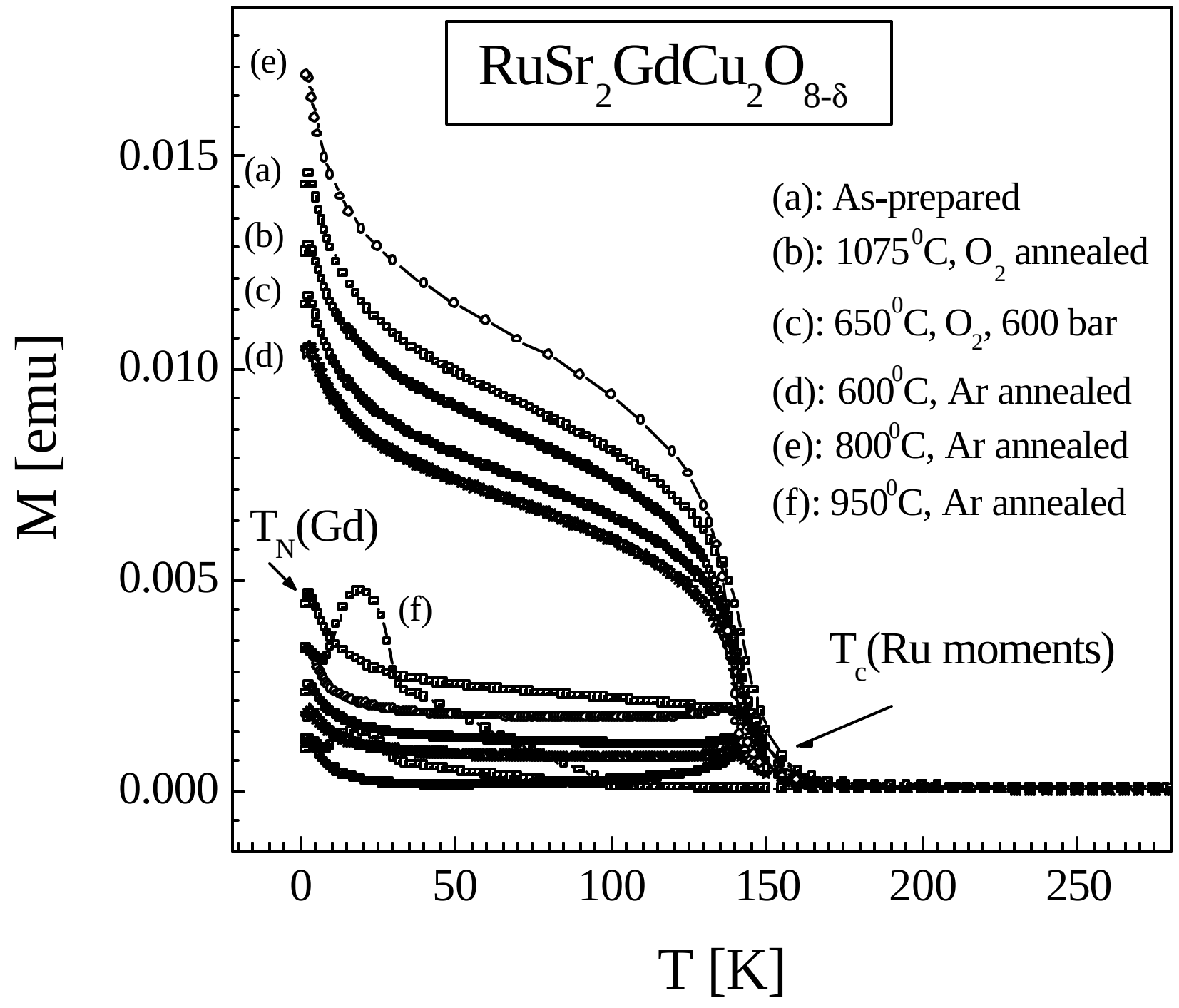


**Fig. 8.** *M*-*T* plots for various non-superconducting $RuSr_2GdCu_2O_{8-\delta}$ samples.

Except the fact that the present samples were non-superconducting and their phase purity was comparatively better than for reported samples, no other visible difference is observed, though it seems that non-superconducting samples have a bit higher magnetic ordering temperature. Worth emphasising is the fact that even after various post-annealing steps, the superconductivity could not be achieved. Also observed is the fact that the magnetic ordering temperatures of Ru or Gd moments were not affected by the post-annealing steps. This highlights the fact, discussed in section 3.2, that there is not much room left for tuning the oxygen stoichiometry of Ru-1212.

The magnetic susceptibility ($\chi$) of the compound follows the paramagnetic behaviour above the magnetic ordering temperature of Ru moments. Considering Gd to be in trivalent state with a localised moment of 8 $\mu_B$ (same as in Gd-based Cu-1212), the calculated moment from Curie-Weiss relation for Ru in paramagnetic state is around 1.0 $\mu_B$, suggesting Ru to be in pentavalent state in Ru-1212. However we should like to mention that moment extraction from Curie-Weiss relation can not be conclusive, without properly considering the exact state of Cu and the effect of possible crystal fields on the magnetic susceptibility of the compound. In fact effective paramagnetic moment for Ru in Ru-1212 has been reported as high as 3.17(4) $\mu_B$ based on high temperature (up to 900 K) fitting of the magnetic susceptibility [12]. The valence of Ru extracted from the magnetic susceptibility data vary from one report to another. We believe the fitting of high temperature magnetic susceptibility to simple Curie-Weiss relation is futile in determining the valence of Ru in rutheno-cuprates like Ru-1212 and Ru-1222, but spectroscopic methods such as XANES are more conclusive in determining the valence state of Ru.

## B. Superconducting Ru-1212 samples

Figure 9 shows the $\chi$-$T$ behaviour in the temperature range of 5 K - 160 K for another as-synthesized $RuSr_2GdCu_2O_{8-\delta}$ sample with an applied field of 5 Oe, in both ZFC and FC situations. The ZFC and FC curves start branching around 140 K with a cusp at 135 K and a diamagnetic transition around 20 K in the ZFC part. The down-turn cusp at 135 K is indicative of antiferromagnetic nature of Ru-spin ordering. Interestingly for the same sample annealed in 100-atm $O_2$ atmosphere the diamagnetic transition was not observed down to 5 K (curve not shown).

For the as-synthesized sample the FC part is seen increasing and later saturating probably due to contribution from paramagnetic Gd moments. Inset of Fig. 9 shows the isothermal *M versus* applied field (*H*) behaviour for this sample. The isothermal magnetization as a function of magnetic field may be viewed as:

$$M(H) = \chi H + \sigma_s(H) , \qquad (1)$$

where $\chi H$ is the linear contribution from antiferromagnetic Ru spins and paramagnetic Gd spins and $\sigma_s(H)$ represents the weak ferromagnetic component of the Ru sublattice. The contribution from the weak ferromagnetic component starts to appear only below 100 K and at higher fields above 3 T. Above this temperature the *M-H* plot remains purely linear. Appearance of ferromagnetic component at low *T* within antiferromagnetically ordered Ru spins can happen due to slight canting of spins. Published neutron diffraction data clearly indicate such a possibility [6-9]. Non-linearity in *M-H* appears at high fields above 3 T. The *M-H* loop for the sample is shown in the lower inset of Fig. 9.

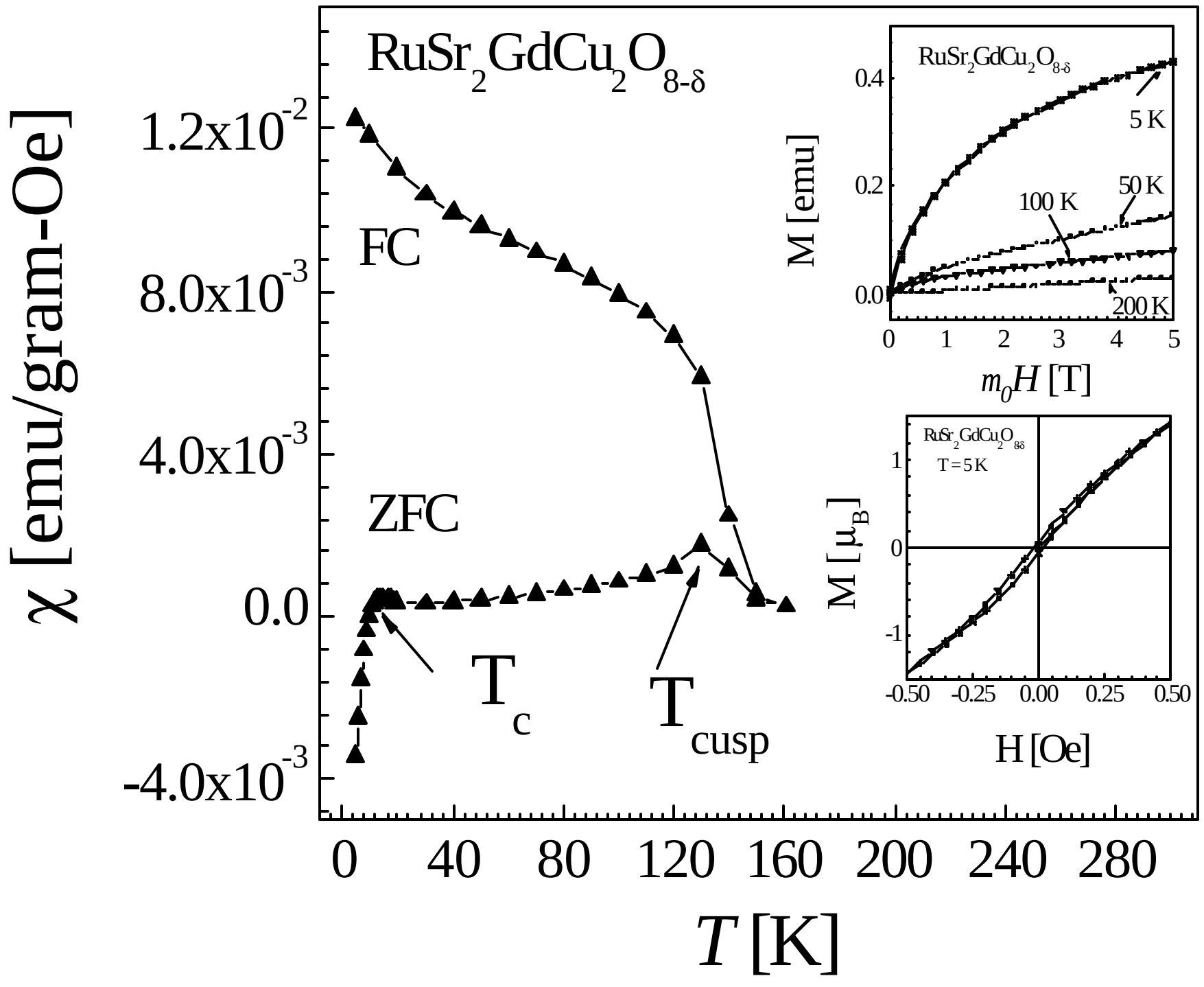

**Fig. 9.** $\chi$-*T* plot for an as-synthesized, superconducting $RuSr_2GdCu_2O_{8-\delta}$ sample. The insets show the *M-H* plots for the same.

### C. Ru-1222 samples

Figure 10 shows the magnetic susceptibility $\chi$-*T* behaviour in the temperature range of 5 to 160 K for an as-synthesized $RuSr_2(Gd_{0.75}Ce_{0.25})_2Cu_2O_{10-\delta}$ sample under applied fields of 5, 10 and 100 Oe, measured in both ZFC and FC modes.

In an applied field of 5 Oe, the ZFC and FC curves start branching at around 140 K with a sharp upward turn at 100 K. The ZFC branch shows further a cusp at 85 K and a diamagnetic transition around 30 K. This is in agreement with earlier reports [1,20,21,23].

The ZFC curve does not show any diamagnetic transition in applied fields of 10 and 100 Oe, but the transition is marked with a change in the slope of the ZFC curves. As the field strength exceeds a certain threshold value the positive contribution from both Gd and Ru moments overcomes the negative contribution from superconductivity to the magnetic susceptibility. Interestingly the ZFC - FC branching temperature of 140 K in 5 Oe field decreases to around 60 K in an applied field of 100 Oe. This can be considered as a weak ferromagnetic behaviour. In fact no ZFC - FC branching is observed down to 5 K in 1,000 and 10,000 Oe fields where both the anomaly and the irreversibility in ZFC and FC branches look to be washed out, see inset in Fig. 10. The down-turn cusp at 85 K in low fields is indicative of antiferromagnetic or spin-glass nature of Ru spins. The FC curve is seen increasing or saturating due to the contribution from paramagnetic Gd spins.

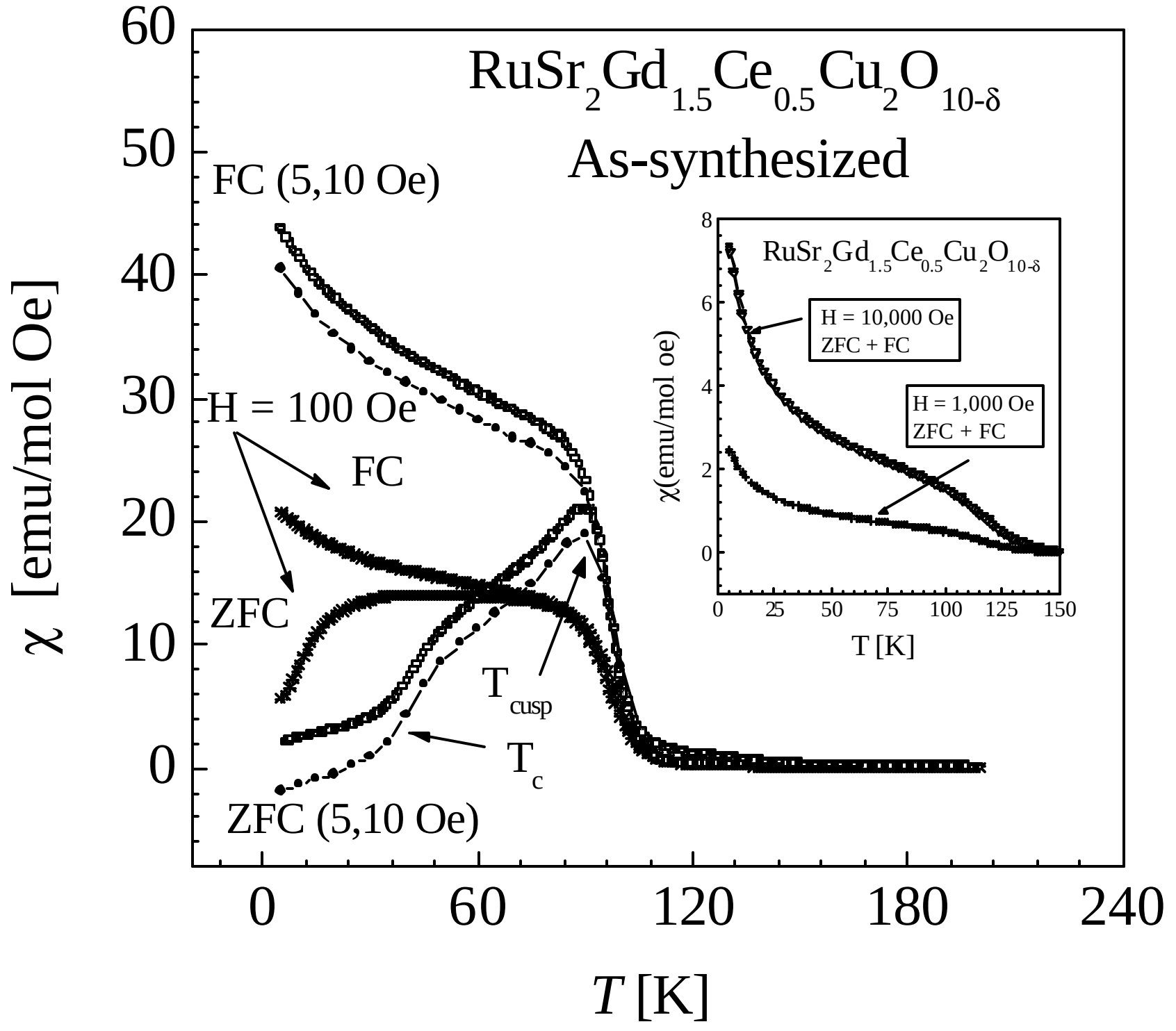

**Fig. 10.** $\chi$-$T$ plot for an as-synthesized $RuSr_2(Gd_{0.75}Ce_{0.25})_2Cu_2O_{10-\delta}$ sample. The inset shows the same at higher fields.

To further elucidate the magnetic property of as-synthesised Ru-1222 we show its isothermal *M-H* behaviour at various temperatures in Fig. 11. The isothermal magnetization as a function of magnetic field may be viewed as previously given equation (1), where $\chi H$ is the linear contribution from antiferromagnetic or spin-glass Ru spins and also from paramagnetic Gd spins, and $\sigma_s(H)$ represents the ferromagnetic component of Ru spins. The appearance of the ferromagnetic component at low temperatures within antiferromagnetic/spin-glass Ru spins could happen due to a slight canting of the spins, as seen from neutron diffraction data for another similar magneto-superconducting Ru-1212 [6-9]. The contribution from the weak ferromagnetic component starts to appear below 150 K. At this temperature the *M-H* plot remains purely linear. The saturation of the isothermal moment appears to occur above say 5 Tesla applied fields, not seen in the present figure. The contribution from the ferromagnetic component starts to appear below 100 K.

The presence of the ferromagnetic component is confirmed by a hysteresis loop measured at 5 K in the *M-H* plot, see inset in Fig. 11. To compare the *M-H* results for Ru-1222 with those for Ru-1212, we showed the same for Ru-1212 in lower inset of Fig. 9. The opening of the *M-H* loop for Ru-1212 is seen in magnetic fields up to 4000 Oe, which is larger in comparison to the 1,000 Oe observed for Ru-1222.

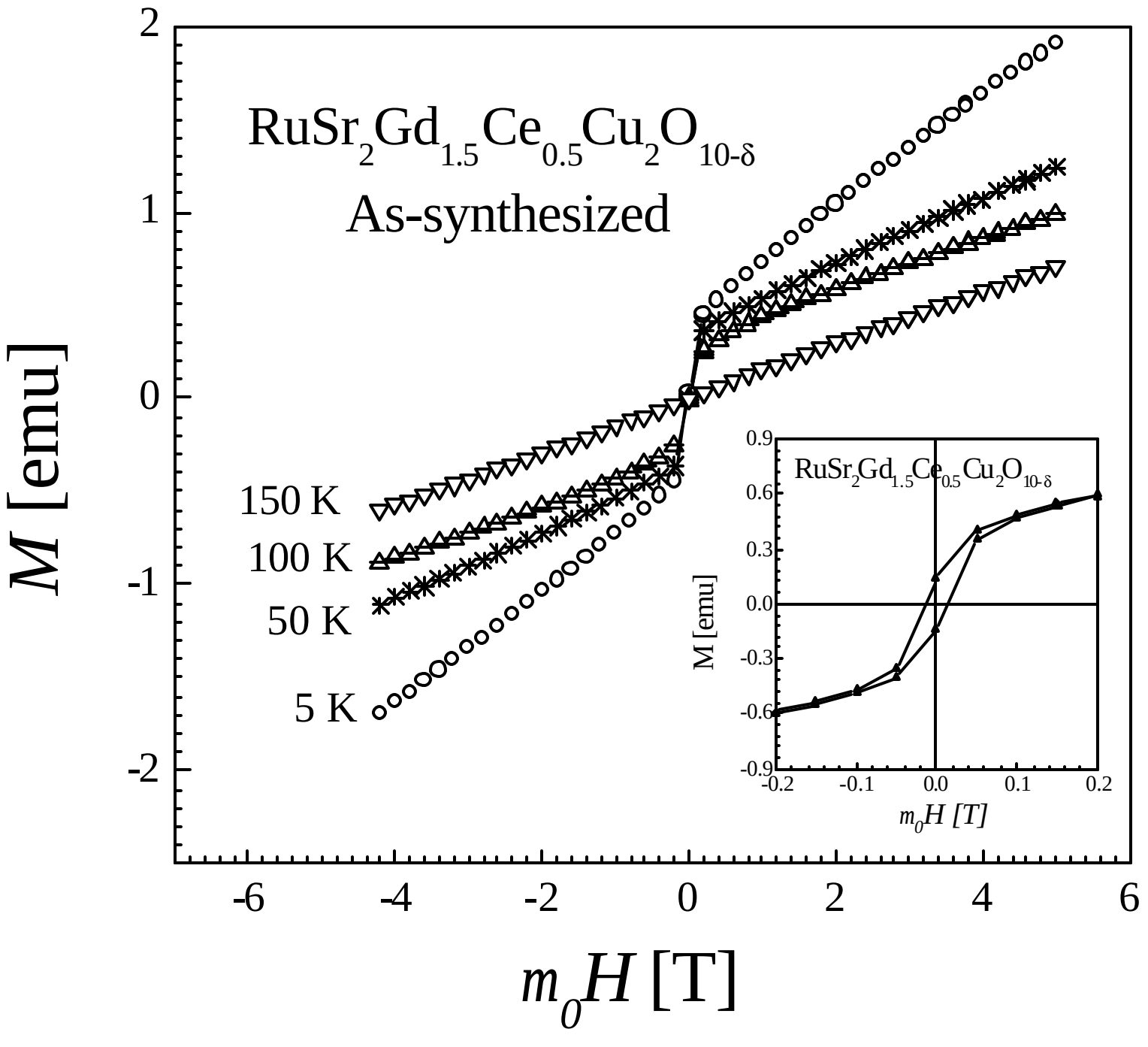

**Fig. 11**. *M-H* behaviour for an as-synthesized $RuSr_2(Gd_{0.75}Ce_{0.25})_2Cu_2O_{10-\delta}$ sample. The inset shows the low field *M-H* hysteresis for the same.

Figure 12 shows the $\chi$-*T* behaviour in the temperature range of 5 to 150 K for the 100-atm $O_2$-annealed sample under applied field of 5 Oe, measured in both ZFC and FC modes. The ZFC and FC curves start branching at around 120 K with a sharp upward turn at 85 K. The ZFC branch shows further a cusp at round 65 K and a diamagnetic transition around 40 K. Also seen is a dip in FC part of magnetization at 40 K, *i.e*, the same temperature where superconductivity sets in. This indicates towards bulk nature of superconductivity in the 100-atm $O_2$-annealed sample.

Interestingly when compared with the as-synthesized sample, one finds that for 100-atm $O_2$-annealed sample while the superconductivity is enhanced, the magnetic ordering (ZFC-FC branching and cusp) temperatures are decreased. Also, the dip in FC magnetization is not seen for the as-synthesized sample. This indicates the competing nature of magnetism and superconductivity in Ru-1222. Weak ferromagnetic component is seen in the 100-atm $O_2$-annealed sample also with nearly same non-linearity field. Summarily, it has been possible to enhance the superconductivity in Ru-1222 by 100-atm $O_2$-annealing.

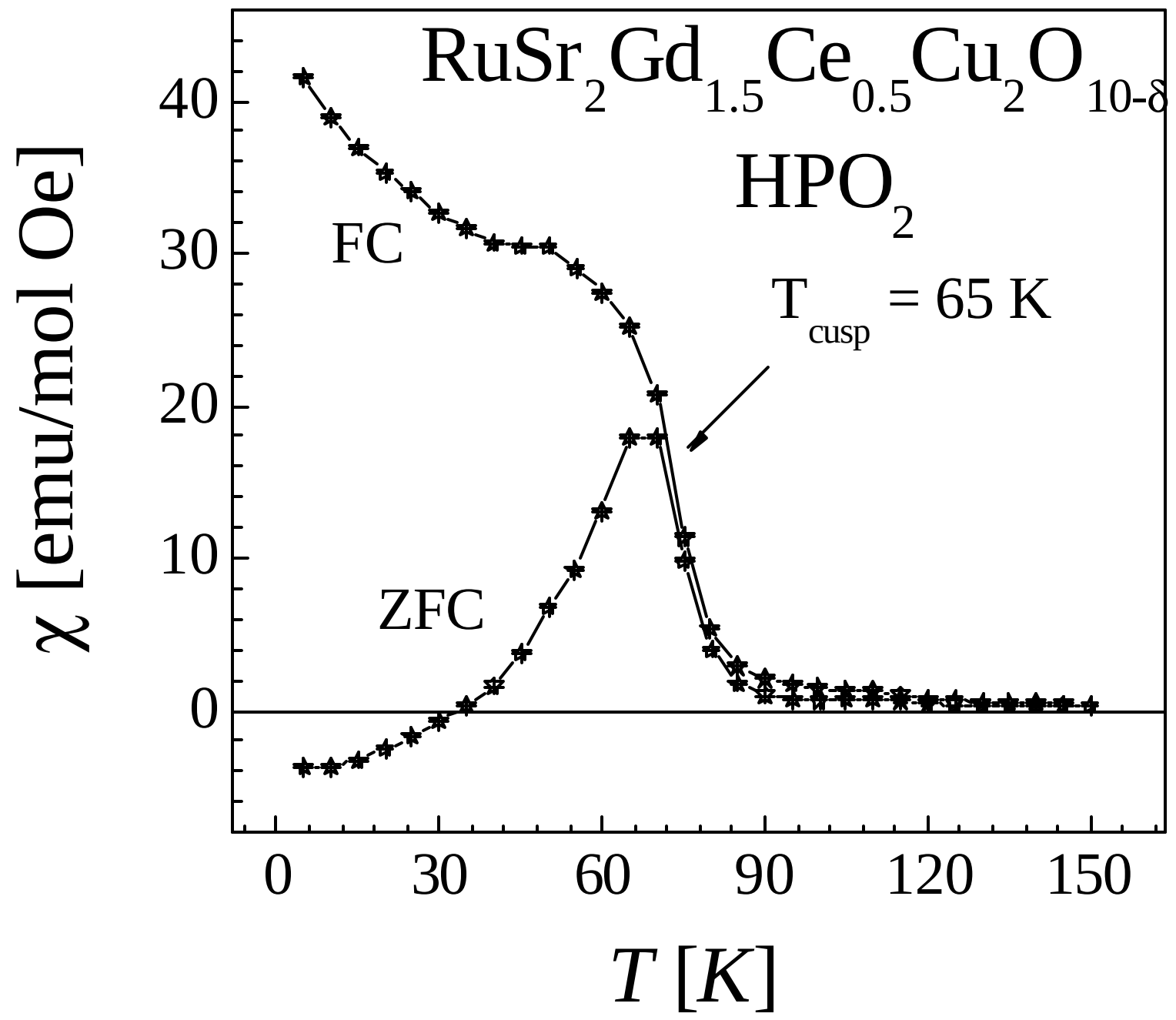

**Fig. 12.** $\chi$-$T$ plot for a 100-atm $O_2$-annealed $RuSr_2(Gd_{0.75}Ce_{0.25})_2Cu_2O_{10-\delta}$ sample.

### 3.6 Electrical transport properties: PPMS results

**A. (Superconducting) Ru-1212 samples**

Figure 13 shows the resistance ($R$) *versus* $T$ for an as-synthesized (superconducting) $RuSr_2GdCu_2O_{8-\delta}$ sample in zero, 3 and 7 T applied fields. The $R$-$T$ behaviour without any applied magnetic field is metallic down to 150 K and semiconducting between 150 K and 25 K, with a superconducting transition onset ($T_c^{onset}$) at 25 K and $R$ = 0 at 20 K. This behaviour is typical of underdoped HTSC compounds. Also observed is an upward hump ($T_{hump}$) in $R$-$T$ around 140 K, which indicates the possibility of antiferromagnetic ordering of spins. The $R$-$T$ behaviour under an applied field of 7 T is nearly the same above $T_c^{onset}$, except that $T_{hump}$ is completely smeared out due to possible change in the magnetic structure. Also in 7 T applied field the $T_c^{onset}$ decreases to around 10 K and $R$ = 0 is not observed down to 5 K.

In an intermediate field of 3 T, both $T_c^{onset}$ and $T_c^{R=0}$ decreased, to 20 and 10 K respectively. For conventional HTSC, $T_c^{onset}$ remains nearly the same under all possible applied fields, with decreasing $R$ = 0 temperature and an increased transition width ($T_c^{onset} - T_c^{R=0}$). Therefore, a different type of broadening of the transition under a magnetic field is obtained for Ru-1212 from that reported for conventional HTSC. In earlier reports on Ru-1212, the $T_c^{onset}$ under a magnetic field decreased like the present case [10,11,14,17,25]. The present behaviour of transition broadening under a magnetic field is presumably due to formation of SNS/SIS junctions/clusters in the present and similar samples. Non-superconducting $RuSr_2GdCu_2O_{8-\delta}$ might be stacked between superconducting $Ru_{1-x}Cu_xSr_2GdCu_2O_{8-\delta}$ resulting in ideal SIS or SNS junctions within the material.

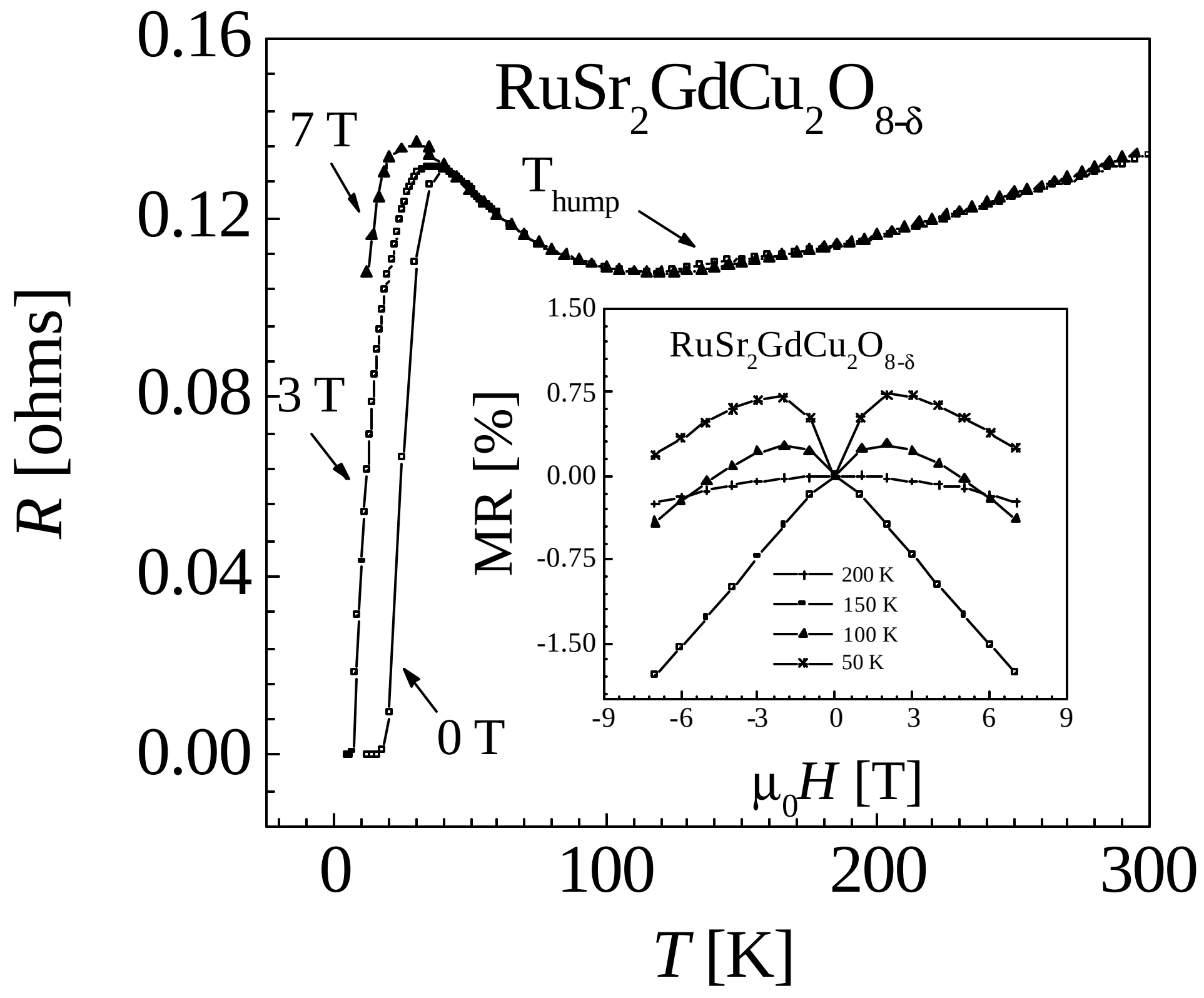

**Fig. 13**. *R-T* plot and magneto-transport behaviour for an as-synthesized (superconducting) $RuSr_2GdCu_2O_{8-\delta}$ sample.

Inset of Fig. 13 shows the magnetoresistance (MR) behaviour of the present Ru-1212 sample at various fields and temperatures. MR is negative in all applied fields upto 7 T above the magne tic ordering temperature, *i.e.* at 150 K and 200 K. Maximum negative MR of up to 2 % is observed at 150 K, which is close to the magnetic ordering temperature of around 140 K. At temperatures below the ordering temperature (100 K and 50 K), MR displays a positive peak at low fields and becomes negative at higher fields. This behaviour is in general agreement with previous reports [10,11,17].

Figure14 shows the $R$-$T$ plot and magneto-transport behaviour for a 100-atm $O_2$-annealed Ru-1212 sample. Interestingly no $R = 0$ is observed only a $T_c^{onset}$ is observed around 20 K. Other characteristics in terms of $T_{hump}$ in $R$-$T$ at around 140 K and the systematic changes in MR with applied field and $T$ are the same as for the as-synthesized sample.

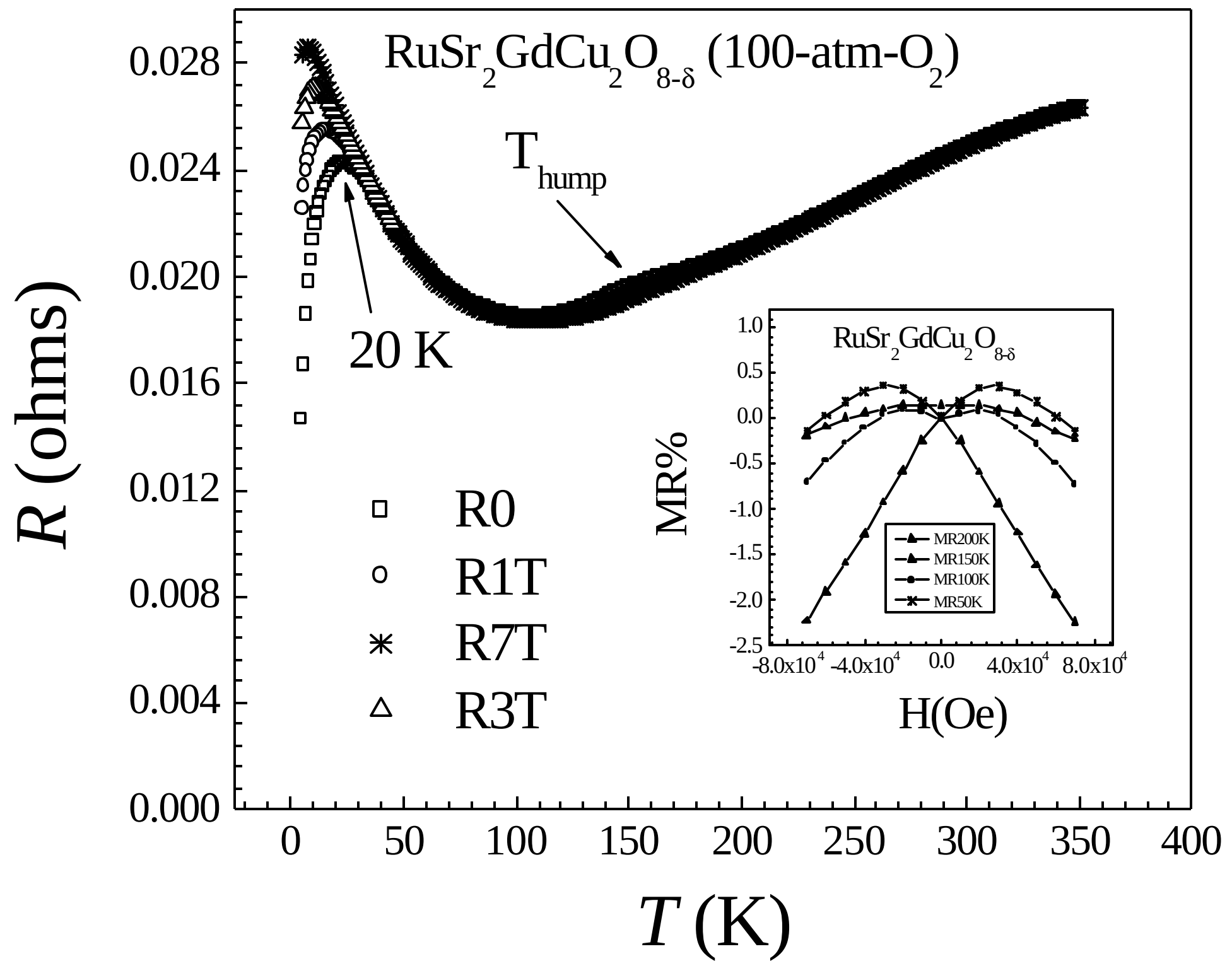

**Fig.14.** $R$ - $T$ and magneto-transport for 100-atm $O_2$-annealed Ru-1212 sample. The inset shows the MR behaviour of the same.

### B. Ru-1222 samples

Figure 15 shows the $R$-$T$ behaviour in zero and 7 T fields for as-synthesized Ru-1222. The $R$-$T$ behaviour in zero field is metallic down to 270 K and semiconducting below 270 K until superconductivity starts with $T_c^{onset}$ at 30 K and $T_c^{R=0}$ at 23 K. The $R$-$T$ behaviour under an applied field of 7 T is nearly the same as that of 0 T above $T_c^{onset}$. However in 7 T field $T_c^{R=0}$ is not observed down to 5 K. In an intermediate field of 3 T, $T_c^{onset}$ and $T_c^{R=0}$ were seen at 25 and 15 K, respectively. Also seen is a shoulder in the $R$-$T$ curve in 3 T, the origin of which is not known.

In the inset of Fig. 15, the MR data of the same Ru-1222 sample is shown at various temperatures and fields, revealing a small negative MR effect in the whole temperature range. Below 100 K the degree of MR is nearly the same in all applied fields and the nature of the MR effect is of the tunnelling-magneto-resistance (TMR) type as judged from the curve shape. Also note that the MR behaviour of the present Ru-1222 sample is different from that of Ru-1212, section 3.6 (A). Ru-1212 exhibited systematic changes in sign of MR at various *T* and fields [10,11,17].

Figure 16 depicts the *R*-*T* plots in 0 and 7 T fields for an 100-atm $O_2$-annealed Ru-1212 sample. The *R*-*T* behaviour in zero field is similar to that observed for the as-synthesized sample with some improvement towards metallic conductivity. Superconductivity starts with $T_c^{onset}$ at 51 K and the $T_c^{R=0}$ is seen at 43 K.

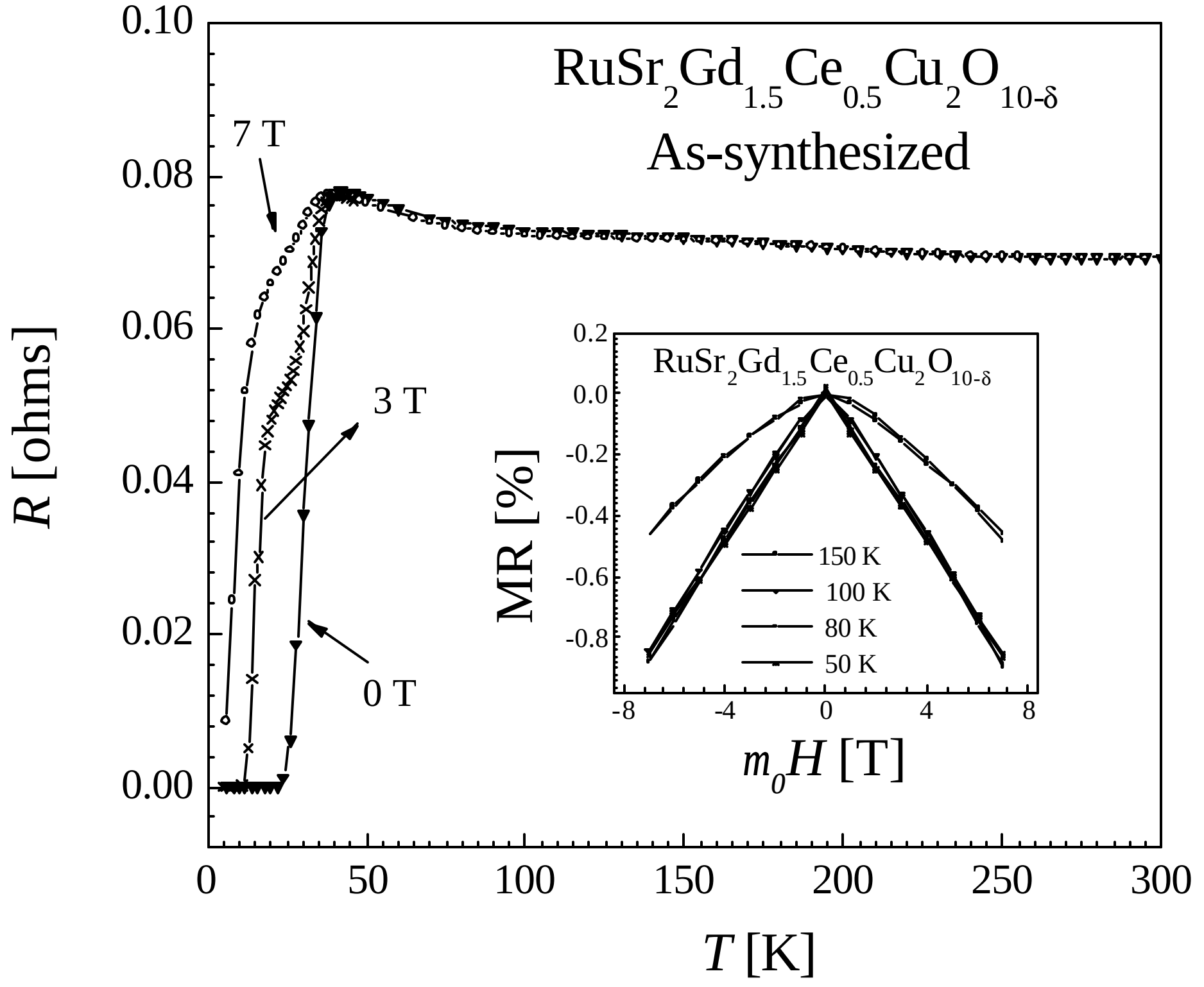

**Fig. 15.** *R*-*T* plot and magneto-transport behaviour for an as-synthesized $RuSr_2(Gd_{0.75}Ce_{0.25})_2Cu_2O_{10-\delta}$ sample. The inset shows the Mr behaviour of the same.

The *R*-*T* behaviour under an applied field of 7 T is nearly the same as that of 0 T above $T_c^{onset}$. However in 7 T field $T_c^{R=0}$ is observed only around 12 K. In intermediate fields of 1 T, 3 T and 5 T, $T_c^{onset}$ remains nearly invariant and $T_c^{R=0}$ are seen at 18 K 16 K, and 14 K, respectively. Also seen is a shoulder in the *R*-*T* curve in all applied fields, the origin of which is not known. When these results of magneto-resistivity are compared with results for the as-synthesized sample (Fig. 11), one finds that $T_c^{R=0}$ is nearly doubled from 23 K to 43 K after the 100-atm $O_2$ annealing. Furthermore, $T_c^{R=0}$ is observed in 7 T field also, which is not the case for the as-synthesized sample. Magneto-resistivity results for the 100-atm $O_2$-annealed Ru-1212 sample substantiate the magnetization results, indicating that superconductivity is enhanced upon the 100-atm $O_2$ annealing. The 100-atm $O_2$-annealed Ru-1222 sample revealed a small negative MR effect (inset in Fig. 16) in the whole temperature range similar to that as for the as-synthesized sample.

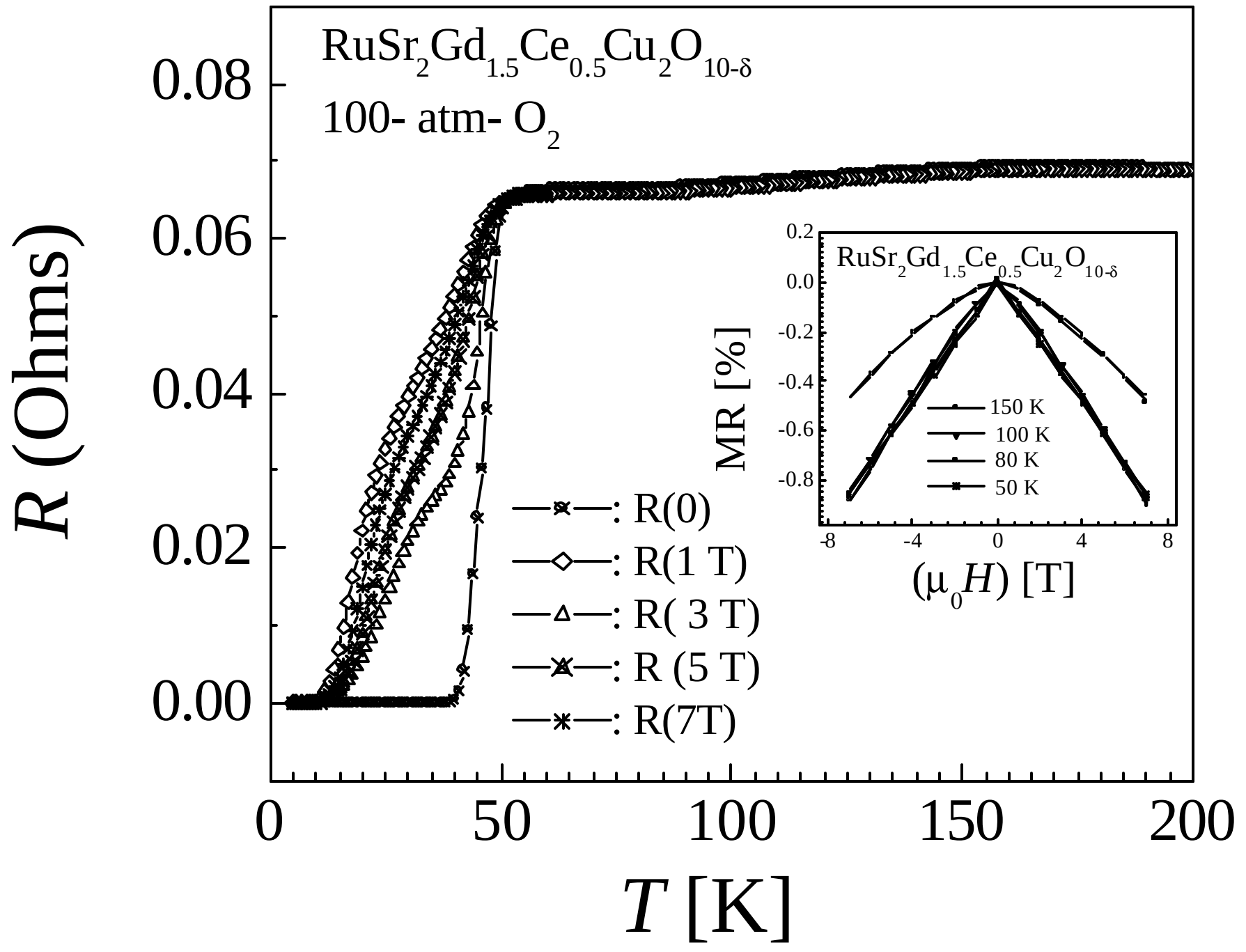

**Fig. 16.** *R*-*T* plot and magneto-transport behaviour for a 100-atm $O_2$-annealed $RuSr_2(Gd_{0.75}Ce_{0.25})_2Cu_2O_{10-\delta}$ sample. The inset shows the MR behaviour of the same.

## 4. SUMMARY AND CONCLUDING REMARKS

We reviewed our various results on rutheno-cuprate magneto-superconductors $RuSr_2GdCu_2O_{8-\delta}$ (Ru-1212) and $RuSr_2(Gd_{0.75}Ce_{0.25})_2Cu_2O_{10-\delta}$ (Ru-1222). It is observed, that it is difficult to control the oxygen content of Ru-1212, though the same is possible up to some extent for Ru-1222. Samples of both phases exhibited superconductivity, presumably in the $CuO_2$ plane at low temperatures coexisting with magnetic ordering of Ru spins of predominantly antiferromagnetic type below 140 K having a ferromagnetic component appearing below 20 K. Electrical conductivity measurements indicate that the $RuO_{2-\delta}$ layer takes part in conduction besides the $CuO_2$ plane. The magnetic ordering temperature of Ru spins is seen as a clear hump in the resistivity measurements, establishing the magnetic spins interaction with the conduction carriers. Though the bulk physical property measurements exhibited the magneto-superconductivity, SAED results question the same to co-existence of the two at microscopic level. Our results indicate towards the fact that solid solutions of $(Ru_{1-x}Cu_x)$-1212, which can be superconducting with $x > 0.5$ [24], but not magnetic may be precipitating with the stoichiometric magnetic Ru-1212 phase. Also highlighted in the review are various existing contradictions in the literature regarding physical properties of these systems. One such example is the results for $C_p$ [16,17] that are used as the evidence of bulk superconductivity in Ru-1212. We believe the establishment of superconductivity along with low-$T$ ferromagnetic ordering of Ru spins in intrinsically phase-pure rutheno-cuprates is yet far from conclusive.